# Vertical-orbital band center as an activity descriptor for hydrogen evolution reaction on single-atom-anchored 2D catalysts

Wen Qiao[1], Shiming Yan[1] [*], Deyou Jin[1], Xiaoyong Xu[2], Wenbo Mi[3], Dunhui Wang[1]

*1 School of Electronics and Information, Hangzhou Dianzi University, Hangzhou 310018, China*

*2 School of Physics Science and Technology, Yangzhou University, Yangzhou, 225002, China*

*3 Tianjin Key Laboratory of Low Dimensional Materials Physics and Preparation Technology,*

*School of Science, Tianjin University, Tianjin 300354, China*

**ABSTRACT:**

The d-band center descriptor based on the adsorption strength of adsorbate has been widely used in understanding and predicting the catalytic activity in various metal catalysts. However, its applicability is unsure for the single-atom-anchored two-dimensional (2D) catalysts. Here, taking the hydrogen (H) adsorption on the single-atom-anchored 2D basal plane as example, we examine the influence of orbitals interaction on the bond strength of hydrogen adsorption. We find that the adsorption of H is formed mainly via the hybridization between the 1s orbital of H and the vertical $d_{z^2}$ orbital of anchored atoms. The other four projected d orbitals ($d_{xy}/d_{x^2-y^2}$, $d_{xz}/d_{yz}$) have no contribution to the H chemical bond. There is an explicit linear relation between the $d_{z^2}$-band center and the H bond strength. The $d_{z^2}$-band center is proposed as an activity descriptor for hydrogen evolution reaction (HER). We demonstrate that the $d_{z^2}$-band center is valid for the single-atom active sites on a single facet, such as the basal plane of 2D nanosheets. For the surface with multiple facets, such as the surface of three-dimensional (3D) polyhedral nanoparticles, the d-band center is more suitable.

[*] Corresponding author: shimingyan@hdu.edu.cn

# 1. Introduction

As a high-density, pollution-free green energy, hydrogen energy is expected to become one of pillar of energy source in the future. Production of hydrogen via electrochemical water splitting is an important way to obtain hydrogen energy, and received greatly attention over the past few decades.[1-4] Finding efficient and inexpensive HER electrocatalyst is the key for the large-scale implementation of water electrolysis technologies.[5-12]

$MoS_2$ is a kind of high-efficiency and cheap 2D transition-metal sulfide (TMD) electrocatalytic material for HER.[13-23] It is considered to be a potential replacement for Pt-based noble metal catalysts. However, its activity is mainly derived from the unsaturated S atoms at the edges, which limit the further enhancement of the HER catalytic activity due to the low atom proportion of the edges. The basal plane with high atom proportion and high exposed area is inert.[23-25] How to add active sites onto the basal plane is one of research subject for such materials in the electrochemical HER. One strategy is to anchor single transition-metal atoms (TMs) on the basal planes. Up to now, many experiment studies have reported highly stable and highly active single-atom-anchored 2D catalyst.[15, 18, 19, 26-30] However, there are few reports associated with the essential mechanism of the activity from the perspective of electronic structure in this new system. Essentially understanding the mechanism of the activity can present rational guiding for the design of high-efficiency catalysts by selecting, high-abundance, low-cost metal atoms and 2D supports as building blocks.

The process of electrocatalytic HER consists of two steps.[31] The first step is the Volmer reaction. In acid media, this step is a process of recombination of electrons and protons. Electrons are transferred to active sites on the catalyst surface and coupled with protons in the solution to generate adsorbed H atoms. The second step is the desorption reaction. This process has two paths, namely the Heyrovsky reaction (the adsorbed H atom and another proton in the solution combine to form H molecule through secondary electron transfer) and the Tafel reaction (two adsorbed H atoms combine to produce H molecule). From the perspective of the entire reaction process, there are two possible paths for the HER, namely Volmer-Heyrovsky and Volmer-Tafel paths. Nevertheless, whichever path, the entire reaction all undergoes the process of H adsorption and desorption. The reaction intermediates of the two paths are all the adsorbed H atoms. It has been verified that an ideal HER electrocatalyst must be facilitated to both H adsorption and desorption.[32, 33] The Gibbs free energy of H adsorption is required to be close to zero, which means that the adsorption strength should be neither too strong nor too weak.[31, 34, 35] So how to control the adsorption strength of H on catalyst is one of the critical parameters in improving the catalytic activity for HER.

D-band center model can well describe the adsorption strength of the adsorbate on metals and their alloys and compounds, particularly in the case of the electronic structure changes under different conditions.[31, 34-38] In terms of this model, trends in the interaction energy between the adsorbate and the metal surface are governed by the coupling to the metal d bands, since the coupling to the metal sp states is essentially the same for the transition and noble metals.[31, 34, 38, 39] The adsorption strength is primarily determined by the energy band position of the d orbital, which is measured by the d-band center. As the d-band center rises, less electrons is filled in d bands, and stronger bond between the metal and adsorbate will be formed. The d-band center descriptor has been widely used to estimate the catalytic activity in various metal alloy catalysts, and given a well guide in the design of high-efficiency catalysts.[36, 37, 40]

However, for the single-atoms active sites on 2D basal planes, not all the d orbitals are effectively hybridized with the H 1s orbital. We know that the d orbital can be subdivided into five projected orbitals $d_{xy}$, $d_{x2-y2}$, $d_{xz}$, $d_{yz}$, and $d_{z2}$. These orbitals have different spatial symmetry and different energy level under the effects of the ligand field. On the 2D basal planes, due to the different symmetry and energy level, these projected d orbitals have different degrees of interaction with the H 1s orbital and thus have different contributions to the binding energy of H adsorption. So the accurate description of HER activity based on the H adsorption energy requires paying more attention to the electronic structure characteristics of the projected d orbitals.

Gibbs free energy calculated by density functional theory (DFT) has been widely used as the reactive descriptor in heterogeneous catalysis. The description of HER activity of single-atom active sites on 2D basal plane also mainly employs this parameter.[16, 18, 22, 41-43] However, this descriptor could not essentially understand the origin of the catalytic activity from the intrinsic features of the catalyst. There are few reports on the mechanism of the activity origin for the single-atom active sites, but the analysis is only based on the energy level of the total d-orbital band.[18, 22, 41] As above mentioned, parts of the projected d orbitals, especially the in-plane projected d orbitals, $d_{xy}$ and $d_{x2-y2}$, are completely inert due to the unique structure of the 2D basal-plane systems. Thus the employing of the total d-orbital band to explore the mechanism in the new system might be sketchy. Up to now, no research focuses on the dependence of features of the projected d orbitals on the activity for the single-atom active sites on 2D basal plane.

Herein, we improve the d-band model and enable it to be more applicable for the single-atom-anchored 2D basal plane system in HER activity. The metallic 1T phase of $MoS_2$ (1T-$MoS_2$) was chosen as the 2D basal-plane substrate. Various TMs were selected to anchor on the basal plane. Using first-principles calculations based on DFT, the interaction between H 1s orbital and the projected d orbitals of the anchored TMs was examined. It is found that the binding of H and the single-atom active sites on basal plane is formed mainly via the hybridization of H 1s orbital and the vertical $d_{z2}$ orbital of the anchored TMs. There is an obvious relationship between the $d_{z2}$-band center and the H bond strength. A descriptor based on the $d_{z2}$-band center is proposed to estimate the electrocatalytic activity of HER. The similarities and differences for the d-band center model and the $d_{z2}$-band center model were discussed. The $d_{z2}$-band center model is the further developing of d-band center model. But they are adapted for different systems. The $d_{z2}$-band center is only valid for the 2D basal plane system, while the d-band center is more suitable for the surface of polyhedral nanoparticles system.

## 2. Methods

We carried the calculations of the energy and electronic structure for all the systems involved using the VASP code. Calculations were performed (with a plane wave basis code) within the generalized gradient approximation, following Perdew-Burke-Ernzerhof (GGA-PBE) for the exchange-correlation function. The plane-wave energy cut-off is 450 eV and an energy convergence threshold is $10^{-6}$ eV. The Monkhorst-Pack k-point mesh of $7 \times 7 \times 1$ was set as the Brillouin-zone integration. All of the TMs anchoring on $MoS_2$ systems were modeled by a supercell of lateral size (2×2) and a vacuum region of 15 Å is used to decouple the periodic images.

The H adsorption energy is estimated as [31]

$$\Delta E_H = E_{\text{2D-BP-SA+H}} - E_{\text{2D-BP-SA}} - \tfrac{1}{2} E_{H2}$$

where $E_{\text{2D-BP-SA}}$ is the total energy for the systems of the TMs-anchored 2D basal plane of 1T-$MoS_2$, $E_{\text{2D-BP-SA+H}}$ is the total energy for these systems with adsorbed H atoms, and the $E_{H2}$ is the energy for a hydrogen molecule in the gas phase.

The H$_2$ molecule adsorption energy is defined as

$$\Delta E_H = E_{\text{2D-BP-SA +H2}} - E_{\text{2D-BP-SA}} - E_{\text{H2}}$$

The d$_{z2}$-band center is calculated with the formula

$$d_{z2} - center = \frac{\int E\rho_{d_{z2}} dE}{\int \rho_{d_{z2}} dE}$$

where, the $\rho_{d_{z2}}$ is the PDOS of the d$_{z2}$ orbital of the TMs at E energy.

The calculation of binding energy is done according to the following formula

$$E_{\text{BE}} = E_{\text{2D-BP+TM}} - E_{\text{2D-BP}} - E_{\text{TM}}$$

where $E_{\text{2D-BP}}$ is the energy of 2D 1T-MoS$_2$ basal plane without anchored TMs, $E_{\text{2D-BP+TM}}$ is the energy of 2D 1T-MoS$_2$ basal plane with anchored TMs, $E_{\text{TM}}$ is the energy of the TMs.

## 3. Results and discussion

The representative system with Co-anchored basal plane of 1T-MoS$_2$ (1T-MoS$_2$-Co) was chosen to examine the feature of H adsorption. On the basal plane of the 1T-MoS$_2$, there are three possible Co-anchored sites, Mo atop site, S atop site and hollow site (as shown in Figure S1). Previous experimental and theoretical studies have shown that the Co atom prefers to bond on the Mo atop sites.[15, 18, 19, 44] Here, the calculated binding energy for Co anchoring also indicates that the Mo atop site is the most stable sites for Co atom (Table S1). This can be ascribed to the extra Mo-Co bonding on the Mo atop site. It is noted that the similar results, due to the same reason, were also found in Fe, Co, Os, Ni anchored 2H phase of MoS$_2$.[26, 29, 45, 46]

For the H adsorption, we consider six kinds of positions (as shown in Figure S1, denoted as numbers) as the initial adsorbed position. After geometric optimization, the six initial geometric configurations all eventually evolve into that with H sitting on the Co atop site (Figure 1a). This suggests that, (1) compared to the S atom, the Co atom is the most attractive atom for H atom; (2) at the Co atop site, the H atom has the most efficiency attractive interaction with Co atom. As discussed below, these results can be attributed to the strong interaction between the unfilled vertical d$_{z2}$ orbital of Co and the 1s orbital of H.

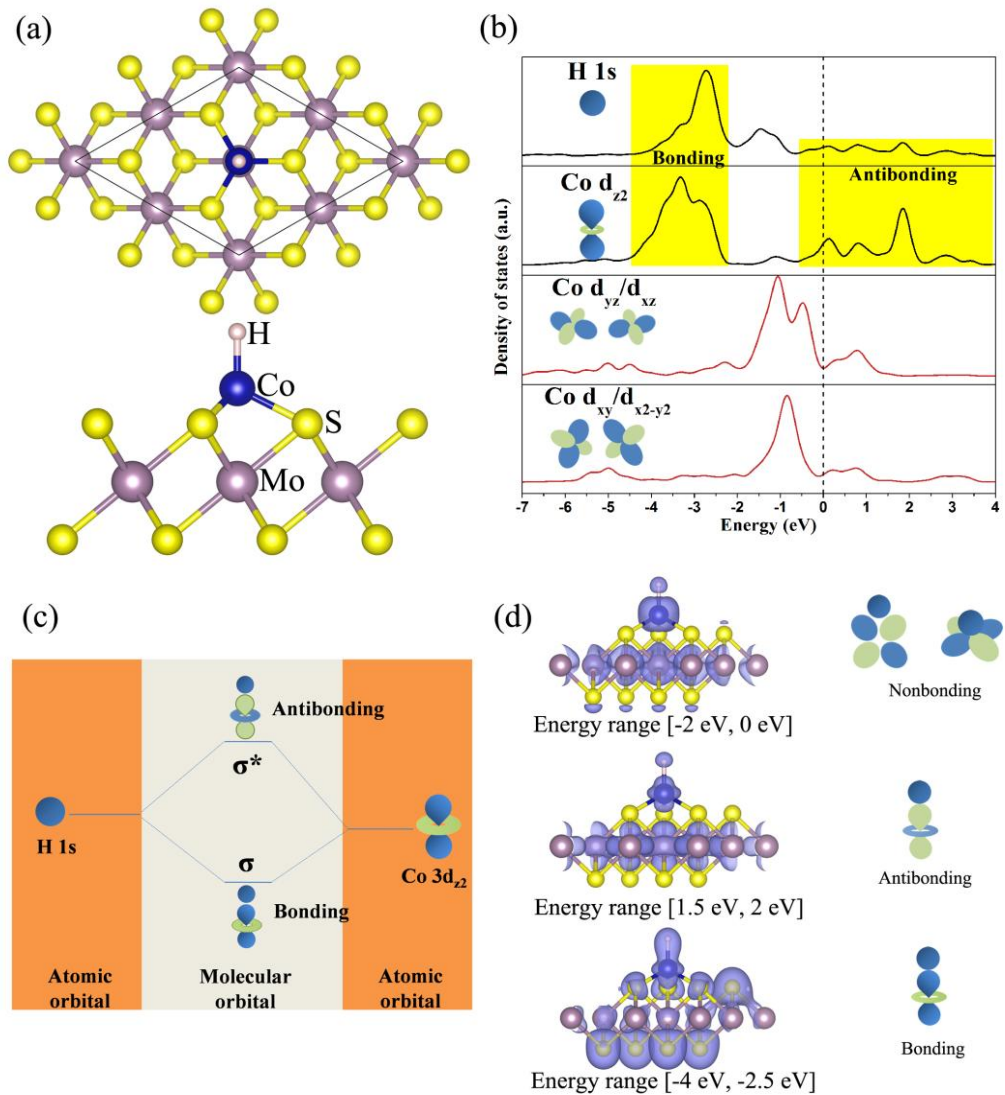

Figure 1. (a) Top and side view of H adsorbed supercell 1T-MoS$_2$-Co after geometric optimization. (b) PDOS of the H 1s, Co d$_{xy}$/d$_{x2-y2}$, Co d$_{xz}$/d$_{yz}$ and Co d$_{z2}$ orbitals for the H-adsorbed 1T-MoS$_2$-Co. (c) Schematic of interaction between H 1s and Co 3d$_{z2}$ orbitals in terms of molecular orbital theory. (d) Charge density of the H-adsorbed 1T-MoS$_2$-Co at different energy range. Right is the schematic illustrations of the corresponding orbitals interaction.

Figure 1b gives the projected density of state (PDOS) of the H 1s, Co d$_{xy}$/d$_{x2-y2}$, Co d$_{xz}$/d$_{yz}$ and Co d$_{z2}$ orbitals for the H-adsorbed 1T-MoS$_2$-Co after geometric optimization. We can see that H 1s orbital profoundly hybridizes with the vertical d$_{z2}$ orbital of the Co atom. Their bonding states locate at the energy range from -4 eV to -2.5 eV. For the corresponding antibonding states, besides a small part below the Fermi energy level, most of them distribute above the Fermi energy level. The interaction between the Co d$_{z2}$ orbital and the H 1s orbital can be understood in terms of molecular orbital theory. As shown in Figure 1c, due to the symmetry matching, along the perpendicular direction of the basal plane, the H 1s orbital overlaps in phase with Co d$_{z2}$ orbital to form σ bonding state at low energy level, and overlaps out of phase to form σ* antibonding state at high energy level. In order to more visibly identify the bonding between H 1s and Co $d_{z2}$ orbitals, we calculated the distribution of charge density of the H-adsorbed 1T-$MoS_2$-Co at different energy range. As shown in Figure 1d, at the energy range from -4 eV to -2.5 eV, where the Co $d_{z2}$ PDOS and H 1s PDOS appear resonance, electrons exhibit integration state at the middle position of H and Co, indicating the bonding state. At the energy range from 1.5 eV to 2 eV, an obvious node lies at the middle position of H and Co, suggesting the antibonding state. At the energy range from -2 eV to 0 eV, where the PDOS of Co is mainly occupied by the $d_{xz}/d_{yz}$ and $d_{x2-y2}/d_{xy}$ orbitals, electrons distribute separately on they own atoms by oneself, meaning no bonding between them. The schematic illustrations of the corresponding orbitals interaction are shown in the right of Figure 1d. For the Co $d_{xz}/d_{yz}$ and $d_{x2-y2}/d_{xy}$ orbitals, the H1s orbital overlap simultaneously with their positive phase and negative phase electron clouds. As a result, no bond is formed.

The hybridization between Co $d_{z2}$ and H 1s orbitals can be identified in the orbital-resolved bands structure. As shown in Figure S2, $d_{x2-y2}/d_{xy}$ bands completely separate from the H 1s orbital bands, indicating no coupling between them; for the $d_{xz}/d_{yz}$ orbitals, there is only a few bands have slightly overlap with the H 1s orbital bands; However, for the $d_{z2}$ orbital, most of its bands profoundly hybridization with the H 1s orbital bands, suggesting their effective bonding.

To further confirm the interaction between Co $d_{z2}$ and H 1s orbitals, we set different distance of Co and H (Co-H) (Figure 2a), and calculated the corresponding PDOS. Decreasing the distance of Co-H will enlarge the overlap and induce high hybridization between Co $d_{z2}$ and H1s orbitals. We can speculate that, according to the molecular orbital theory, this will amplify the energy splitting of the bonding and antibonding states. Figure 2b shows the calculated PDOS. We can see that at large Co-H distance of 4.590 Å, the feature of Co $d_{z2}$ PDOS is almost the same as that without H adsorption (Figure 3a), indicating no interaction between the Co and H at this distance. As the Co-H distance decreases to 2.470 Å, the Co $d_{z2}$ and H 1s orbitals begin to hybridize. The PDOS of H 1s orbital appears downshift and resonance with the PDOS of Co $d_{z2}$ orbital. With further decreasing the distance, the resonance strength become stronger, the bonding state shift to lower energy level, and its energy difference compared to the antibonding state become more and more large. Different form the vertical $d_{z2}$ orbital, there is no changes for the PDOS of $d_{xy}/d_{x2-y2}$ and $d_{xz}/d_{yz}$ orbitals with decreasing Co-H distance (Figure S3). These results further suggest that H

1s orbital only hybridizes with the vertical $d_{z^2}$ orbital at the basal plane. For the nonvertical orbitals, $d_{xy}/d_{x^2-y^2}$ and $d_{xz}/d_{yz}$, the H 1s orbital has no bonding with them.

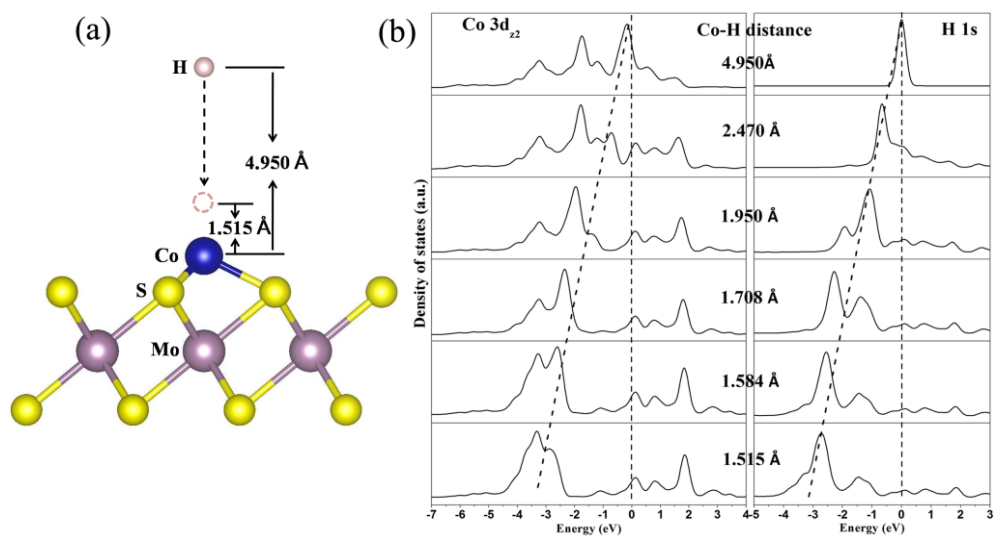

Figure 2. (a) Schematic for altering the Co-H distance. (b) PDOS of Co $3d_{z^2}$ and H 1s orbitals at different distance of Co-H.

Various factors of electric structure can influence the H adsorption energy on catalysis, such as the local d-band density of states at the Fermi level, the number of holes in the d bands, the width and shape of the d bands.[39] However, these factors usually could not measure the adsorption energy solely or are generally smaller corrections.[38] The d-band center has been verified to give a primary contribution towards the surface adsorption reactive of metals.[34-36, 38] For the 2D basal plane systems, since the chemical bond of H forms mainly through the interaction of the vertical $d_{z^2}$ orbital, the band center of $d_{z^2}$ orbital could play a primary role in determining the H adsorption energy. In order to confirm this, we calculated the PDOS of $d_{z^2}$ of different 3d TMs anchored on the basal plane of 1T-$MoS_2$, and examined the correlation between the $d_{z^2}$-band center and H adsorption energy.

Similar to the Co anchoring, compared to the hollow site and S atop site, other 3d TMs also prefer to bond on the Mo atop site (Table S). So here we select this geometric configuration as the object to study the influence of $d_{z^2}$-orbital electronic structure on the strength of H adsorption. Figure 3a shows the $d_{z^2}$ PDOS of the series anchored 3d TMs. The $d_{z^2}$-band center calculated from the $d_{z^2}$ PDOS is shown in Figure 3b. The calculated H adsorption energy on the series 3d TMs is shown in the Figure 3c. We can see that, for the left atoms of the first row TM elements in the periodic table, the $d_{z^2}$-band centers locate at high energy level; while, at right of the periodic table, the $d_{z^2}$-band centers of the 3d TMs locate at low energy level. From right Ni to left V, the $d_{z^2}$-band center shifts up gradually to high energy (Figure 3b). This relation can be attributed to the difference of the 3d orbital filling in the series 3d TMs. For the first row TM elements, the filling of 3d orbital gradually increase from left to right in the periodic table. Increasing the filling of 3d orbital will result in the reduction of the band center. Because the more filling means that more electrons occupy the states below Fermi level. As a consequence, the band center will be low away from Fermi level.

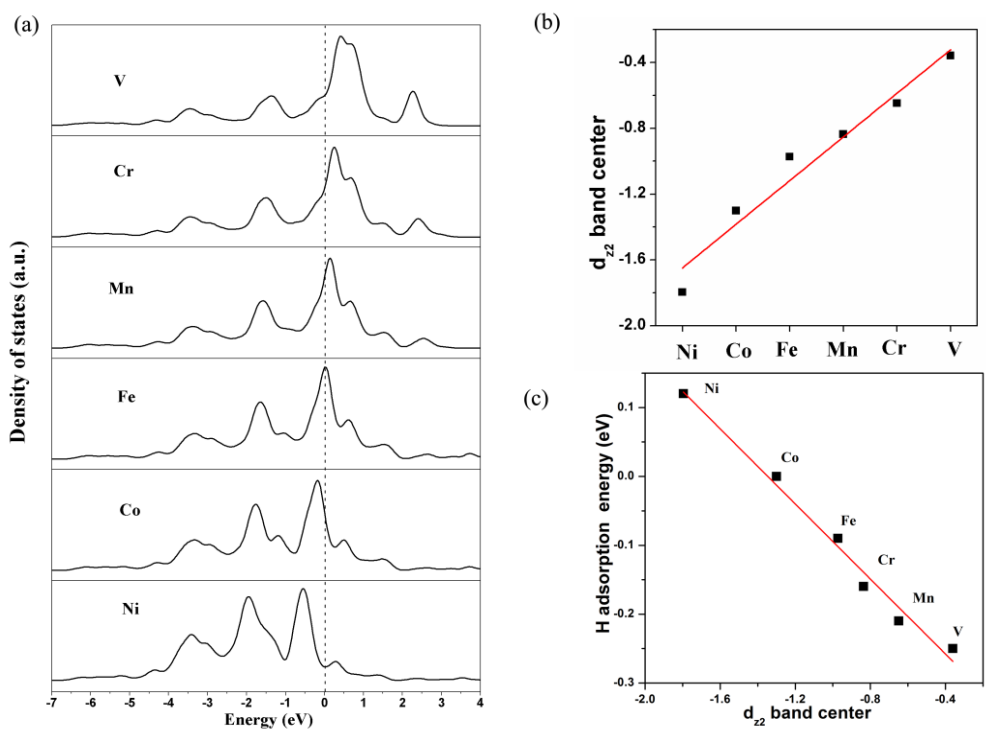

Figure 3. (a) PDOS of $d_{z^2}$ for the series anchored 3d TMs on 1T-MoS$_2$. (b) Caculated $d_{z^2}$-band center for the series 3d TMs on basal planes of 1T-MoS$_2$. (c) The $d_{z^2}$-band center dependence of H adsorption energy for the series 3d TMs anchored on basal planes of 1T-MoS$_2$.

Figure 3c shows the $d_{z^2}$-band center dependence of H adsorption energy. There is obvious correlation between them. With increase of the $d_{z^2}$-band center, the H adsorption energy decreases approximately linearly. This relationship is the result of the different electron filling of the antibonding state formed by the hybridization with H. It is known that the bond strength is determined by the filling of the antibonding state.[34, 35] The more the empty antibonding state is, the higher the strength of bond is. Since the antibonding states are always above the $d_{z^2}$ states, increase of $d_{z^2}$-band center will result in the rise of the antibonding state. Our calculated PDOS of the TMs with H adsorbed indeed show the upshift of the antibonding state with increase of the

$d_{z2}$-band center (Figure S4). Therefore, increasing $d_{z2}$-band center will lead to the strong H bond strength and low H adsorption energy.

Besides the filling of orbital, another way to change the position of d-band center is the strain. Strain has large influence on the width of d bands, but negligible influence on the d orbital filling. This will fix the d band to the Fermi level. With adding strain, system will have to compensate for variations in the width by shifting the d states up or down in energy.[35, 36] In order to further confirm the correlation between the $d_{z2}$-band center and H adsorption energy, we added different strain to the typical system 1T-MoS$_2$-Co, and calculated the PDOS and H adsorption energy. The properties, in particular the energy, associated with $d_{z2}$ orbital are sensitive to the perpendicular strain. As the orientation of the $d_{z2}$ orbital is perpendicular to the basal plane, strain along this direction will significantly influence the overlap of the $d_{z2}$ orbital with others. Therefore, here we select to add the strain along the c axis, instead of a or b axis.

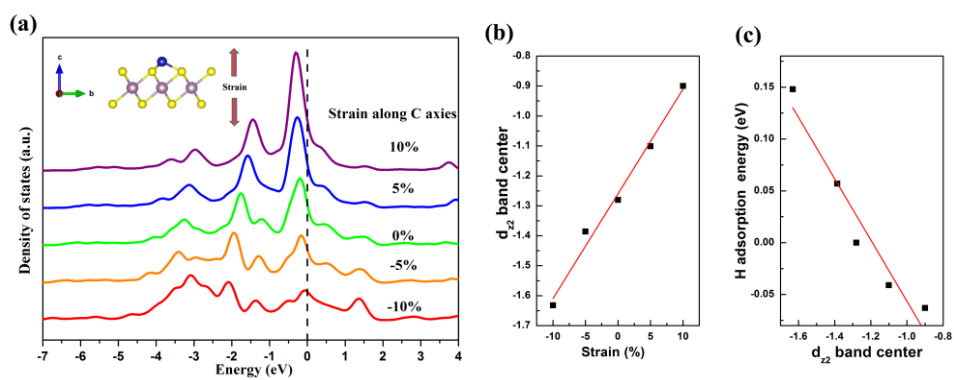

Figure 4. (a) PDOS of $d_{z2}$ orbital under different perpendicular strains for the 1T-MoS$_2$-Co. (b) The strain dependence of $d_{z2}$-band center for the 1T-MoS$_2$-Co. (c) The $d_{z2}$-band center dependence of H adsorption energy for the 1T-MoS$_2$-Co.

Figure 4a gives the PDOS of $d_{z2}$ orbital under different perpendicular strains for the 1T-MoS$_2$-Co. We can see that the tensile strain results in the band narrowing, and simultaneously the PDOS below Fermi level shift up to higher energy; while the compression strain leads to the change on the contrary. The calculated $d_{z2}$-band center show proportionally dependence with variation of strain (Figure 4b). As is expected, the H adsorption energy also shows proportionally relationship with strain. Tensile strain reduces the H adsorption energy; compression strain enhances the H adsorption energy. The correlation between the H adsorption energy and the $d_{z2}$-band center was plotted in Figure 4c. Obviously, with increasing the band center, the H adsorption energy decrease linearly. This further confirms the influence of $d_{z^2}$-band center on the H bond strength.

In addition, we also calculated the adsorption energy of $H_2$ molecule on different 3d TM atoms anchored on 1T-$MoS_2$ (shown in Figure S6). For each 3d TM, the adsorption energy of $H_2$ molecule is larger than the H adsorption energy, indicating that $H_2$ molecule is unstable on the TM atoms compared to the adsorbed H. The forming $H_2$ molecule is easy to desorb from the TM atoms, and do not influence the adsorption of H atom. From the Figure S6, we can see that, similar to the H adsorption, $H_2$ molecule adsorption energy also decrease linearly with increasing the $d_{z^2}$-band center. This result suggests that, for adsorption of H atom and $H_2$ molecule, the binding strengths between H and 3d TM are both determined by the $d_{z^2}$-band center.

Due to the mainly determinant of catalytic activity for HER by H bond strength and the explicit linear relation between the H bond strength and the $d_{z^2}$-band center, we can use the $d_{z^2}$-band center as the descriptor for the active in HER. Similar to the d-band center descriptor, if the band center is too high, the H will bind strongly to the surface and have difficulty leaving from the surface; if band center is too low, the H will bind too weakly and be difficult to adsorb. High catalytic activity for HER needs a modest band center. It should be mentioned that the single atom on 2D surface is not absolute alone. It is anchored by the chemical bond on the 2D surface. The single atom here is a part of the 2D basal plane and acts as an atomic-level reactive site. Its s, p and d bands all hybridize with those of other atoms in the basal plane substrate and show continuous distribution (as shown in Figure S5). Therefore the center of energy bands can be applied to descript the reactive for this system with single-atom active sites. Furthermore, the explicit linearly relation between the band-center and the H bond strength directly confirm this feasibility.

There are both similarities and differences for the d-band center descriptor and the $d_{z^2}$-band center descriptor. The similarity is that the evaluation of the catalytic activity are all based on the efficiency of the adsorbate adsorption/desorption which is dominated by the bond strength of adsorbate on surface. This idea runs through our full text. The distinct difference is the reactive orbital. One is the d orbital and the other is the $d_{z^2}$ orbital. Another difference is that they are suitable for different single-atom-anchored surface systems. The d-band center is suitable for the surface of nanoparticles system; while, the $d_{z^2}$-band center is valid for the surface of 2D basal plane system. In the following, we discuss this implied difference.

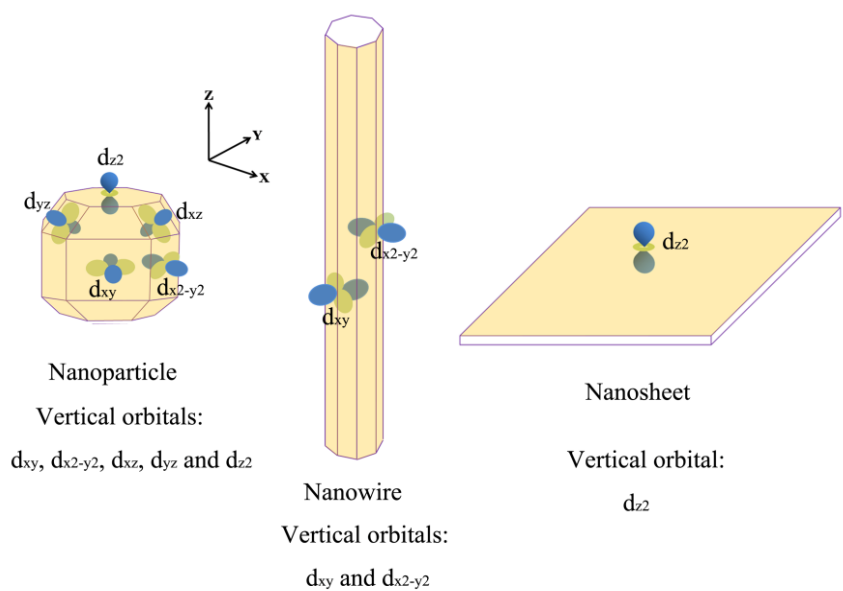

Figure 5. Vertical orbitals at various surfaces in different-dimensional systems, coordinate axis is as shown in the figure.

In catalytic reactive, the adsorbate adsorption on surface of catalysis always involve the interaction between the orbitals of catalyst active site and adsorbate. As above demonstrated, not all the orbitals of the active site participate in the interaction. For the HER of single-atom active sites, only the orbitals perpendicular to the surface can have effective interaction with H 1s orbital. In this way, the strength of the H adsorption on the surface of catalysis is mainly determined by the feature of these vertical orbitals. For a single flat surface, such as the 2D basal plane, there is only one projected d orbital perpendicular to it. If the coordinate Z axis is perpendicular to the surface, this vertical projected d orbital is the $d_{z^2}$ orbital as shown in Figure 5. Differently, for 3D polyhedral systems, such as nanoparticles, the surface is composed of many facets with different direction. Taking the polyhedral nanoparticles with cubic crystal structure as example, the projected d orbitals $d_{z^2}$, $d_{xy}$, $d_{x^2-y^2}$, $d_{xz}$, and $d_{yz}$ are perpendicular separately to five different facets with orientations (001), (110), (100), (101) and (011), respectively (as shown in Figure 5). In this case, the five projected d orbitals from the active sites on the different facets can all hybrid with H 1s orbital, and thus the band center of the each projected d orbital can descript the HER activity on their corresponding facets. The average of the band centers for the five projected d orbitals is equal to the d-band center. So the d-band center can be used to estimate the overall catalytic activity for the surface of polyhedral systems.

However, for the new system of the 2D basal plane, the vertical d orbital from the anchored active TM atom is just the $d_{z^2}$ orbital. Therefore, the $d_{z^2}$-band center is only suitable for such systems.

Based on the above discussion, we can deduce that, the band center of $d_{xy}/d_{x^2-y^2}$ orbitals might be used as the descriptor for HER on the side face of the 1D nanowires or nanorods system. Because, assuming the Z axis is along the direction of the nanowires/nanorods, the vertical orbitals on the side face is the $d_{xy}/d_{x^2-y^2}$ (as shown Figure 5), which is the unique orbitals that can form bond with H atom compared to the other projected d orbitals.

## 4. Conclusion

In conclusion, aiming at the single-atom-anchored 2D catalysts, a new descriptor, $d_{z^2}$-band center for the HER activity were proposed based on the hydrogen adsorption energy. Due to the perpendicular feature to surface for the $d_{z^2}$ orbital, for more extensive significance the descriptor can also be termed as vertical-orbital band center descriptor. Different from the traditional d-band center which can be well used to measure the catalytic activity in nanoparticles system, this descriptor is valid for the 2D basal plane system.

## Acknowledgements

This work was supported by the National Natural Science Foundation of China (Grant No. 11504086), the Ten Thousand Talents Plan of Zhejiang Province of China (Grant No. 2019R52014), the Open Project of National Laboratory of Solid State Microstructures, Nanjing University (Grant No. M33010) and the School Scientific Research Project of Hangzhou Dianzi University (Grant Nos. KYS045619084, KYS045619085).

# Supporting Information

Vertical-orbital band center as an activity descriptor for hydrogen evolution reaction on single-atom-anchored 2D catalysts

Wen Qiao[1], Shiming Yan[1,*], Deyou Jin[1], Xiaoyong Xu[2], Wenbo Mi[3], Dunhui Wang[1]

*1 School of Electronics and Information, Hangzhou Dianzi University, Hangzhou 310018, China*

*2 School of Physics Science and Technology, Yangzhou University, Yangzhou, 225002, China*

*3 Tianjin Key Laboratory of Low Dimensional Materials Physics and Preparation Technology, School of Science, Tianjin University, Tianjin 300354, China*

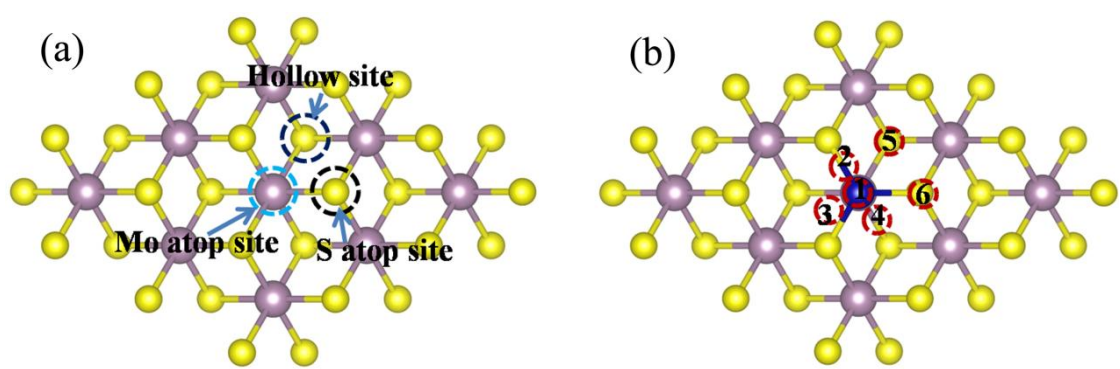

Figure S1. Schematic of possible TM anchoring site. (b) Initial H adsorbed position for geometric optimizing.

Table S1. Binding energy for series TMs occupying on different surface site.

| | Binding Energy (eV) | | | | | |
|---|---|---|---|---|---|---|
| | V | Cr | Mn | Fe | Co | Ni |
| **Mo atop site** | -3.37 | -1.46 | -1.37 | -2.94 | -4.09 | -3.81 |
| **S atop site** | 0.04 | 2.02 | 1.99 | 0.04 | -1.46 | -2.04 |
| **Hollow site** | -3.30 | -1.40 | -1.23 | -2.57 | -3.56 | -3.22 |

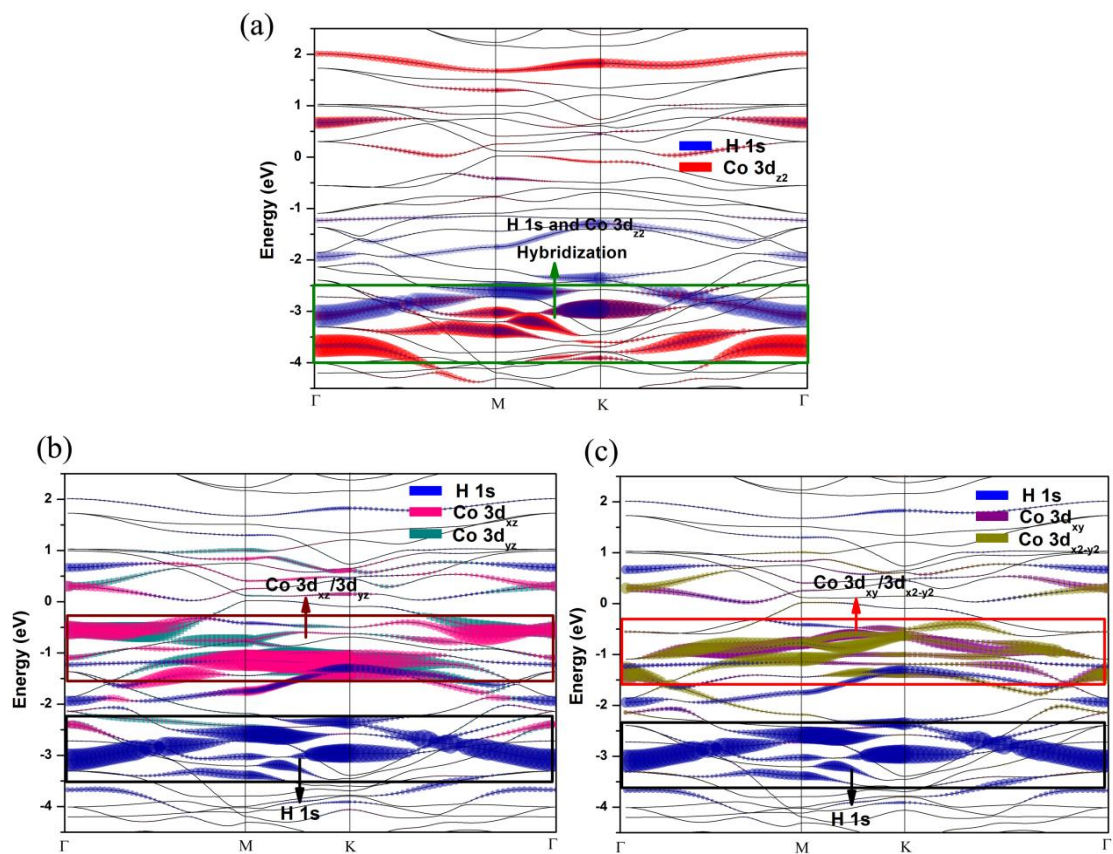

Figure S2. The calculated orbital-resolved bands structure for the 1T-MoS$_2$-Co with H adsorption.

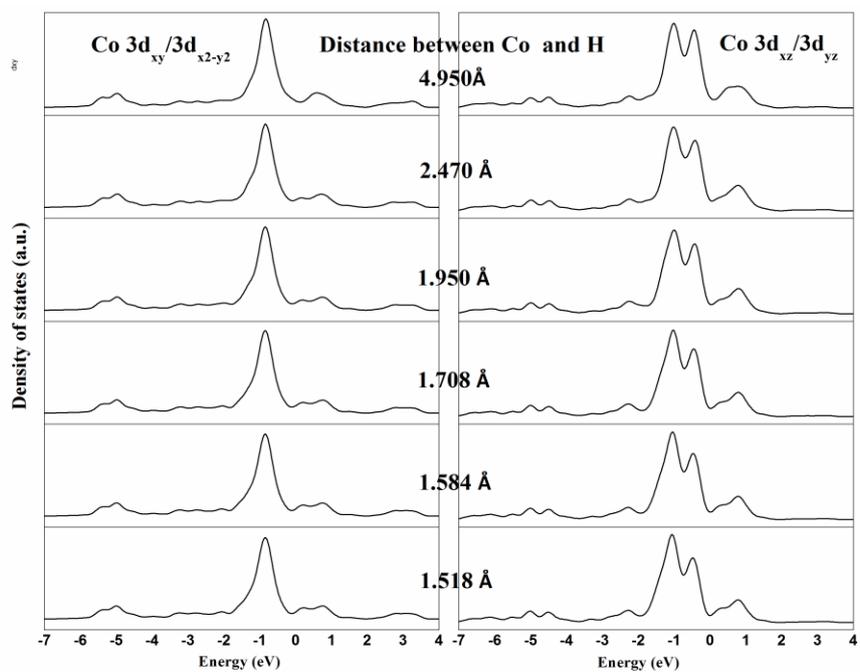

Figure S3. PDOS of Co $3d_{xy}/3d_{x2-y2}$ and $3d_{xz}/3d_{yz}$ orbitals at different the distance of Co-H.

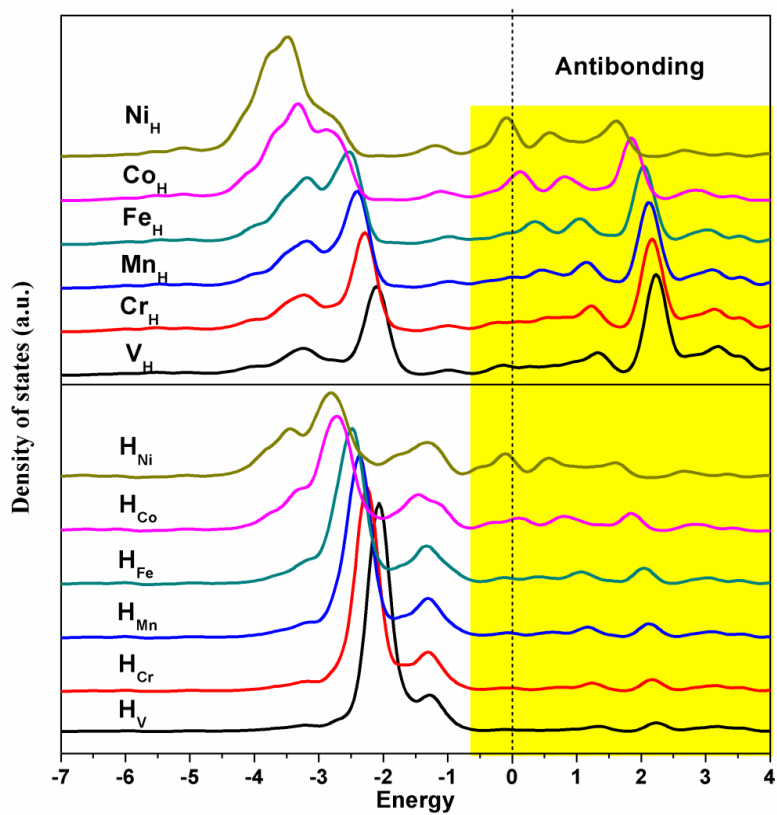

Figure S4. PDOSs of the TMs $d_{z2}$ orbitals and the corresponding H 1s orbitals in series TMs anchoring on 1T-$MoS_2$ with H adsorption.

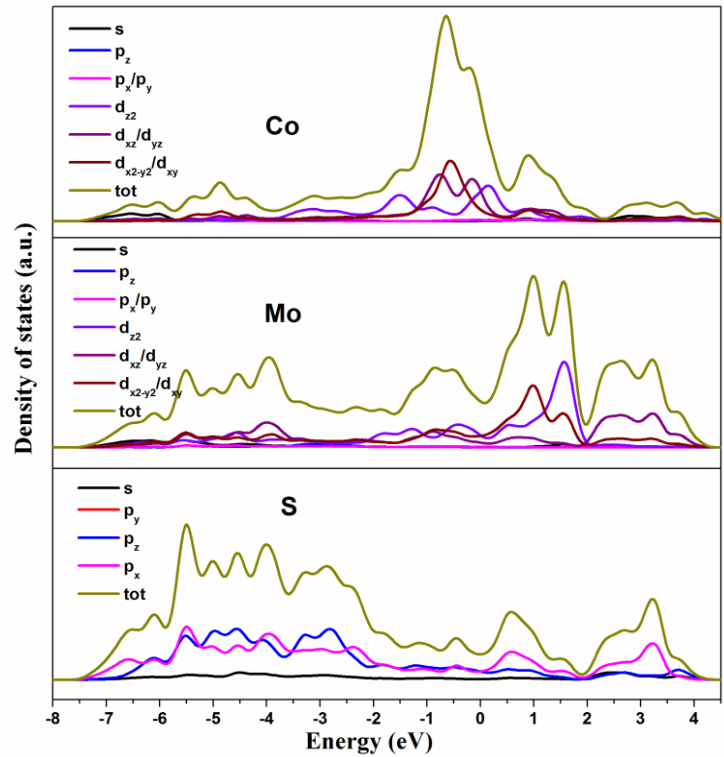

Figure S5. DOSs and PDOSs of the different orbitals for Co, Mo and S atoms in 1T-MoS$_2$-Co.

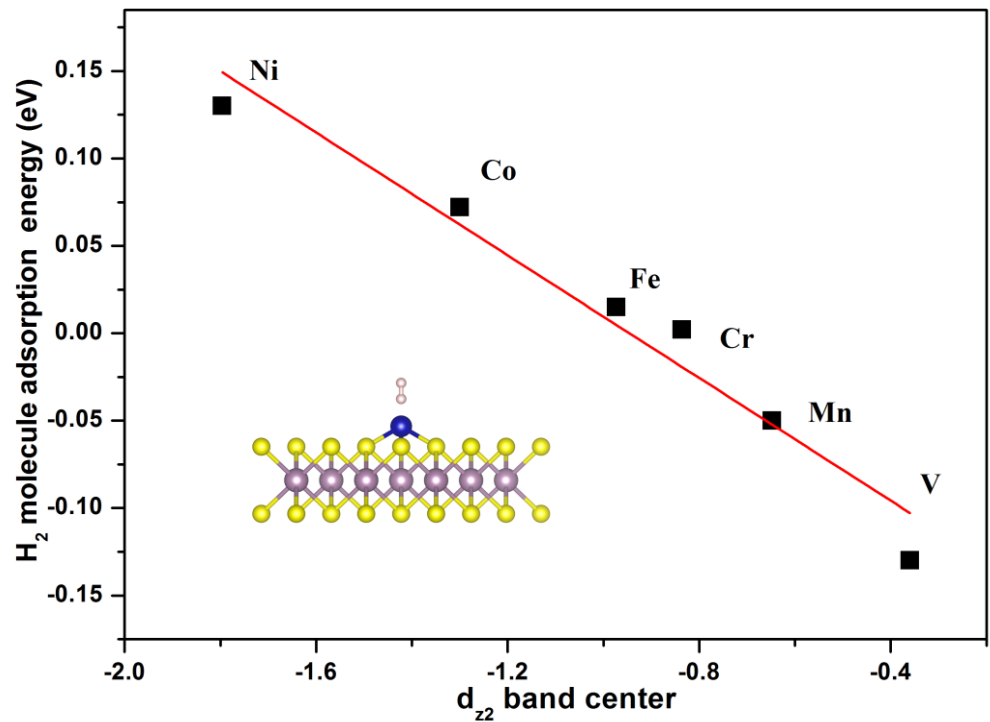

Figure S6. Dependence of the d$_{z2}$-band center on H$_2$ molecule adsorption energy for the series 3d TMs anchored on basal planes of 1T-MoS$_2$.

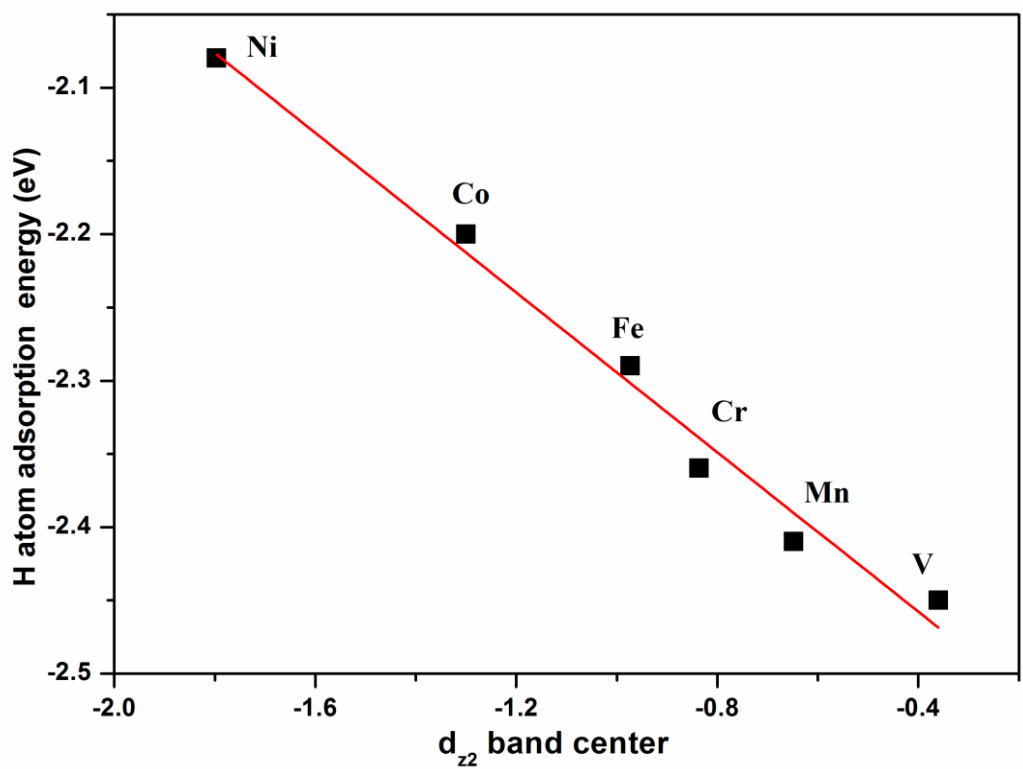

Figure S7. Dependence of $d_{z2}$-band center on H atom adsorption energy for the series 3d TMs anchored on basal planes of 1T-MoS$_2$. Different form H adsorption energy, the H atom adsorption energy is defined as $\Delta E_H = E_{\text{2D-BP-SA +H}} - E_{\text{2D-BP-SA}} - E_H$.